\documentclass[11pt]{article}
\usepackage[utf8]{inputenc}
\usepackage{amsmath,amsfonts,amssymb}
\usepackage{amsthm}
\usepackage{latexsym}
\usepackage[noadjust]{cite}
\usepackage{graphicx}
\usepackage{xcolor}
\usepackage{mathtools}
\usepackage[margin=1in]{geometry}
\usepackage{thm-restate}
\usepackage{enumitem}
\usepackage[ruled,vlined,linesnumbered]{algorithm2e}
\usepackage{tikz}
\usepackage{subcaption}
\usepackage[framemethod=TikZ]{mdframed}
\usepackage{booktabs}

\definecolor{ForestGreen}{rgb}{0.1333,0.5451,0.1333}
\definecolor{DarkRed}{rgb}{0.8,0,0}
\definecolor{Red}{rgb}{1,0,0}
\usepackage[linktocpage=true,
pagebackref=true,colorlinks,
linkcolor=DarkRed,citecolor=ForestGreen,
bookmarks,bookmarksopen,bookmarksnumbered]
{hyperref}

\usepackage[capitalise]{cleveref}

\newtheorem{theorem}{Theorem}[section]
\newtheorem{corollary}[theorem]{Corollary}
\newtheorem{lemma}[theorem]{Lemma}

\newtheorem{claim}[theorem]{Claim}
\theoremstyle{definition}
\newtheorem{definition}[theorem]{Definition}
\newtheorem{remark}[theorem]{Remark}
\newtheorem{fact}[theorem]{Fact}

\newcommand{\E}{\mathbb{E}}
\newcommand{\MSF}{\textnormal{\textsf{MSF}}}
\newcommand{\Cert}{\textnormal{\textsf{Certificate}}}

\title{A Deterministic Linear Time Algorithm for \\ Minimum Spanning Tree}
\date{\today}

\author{Sanjeev Khanna\thanks{Courant Institute, New York University, New York, NY, USA.  Supported in part by NSF award CCF-2625203 and AFOSR award FA9550-25-1-0107. Email: {\tt sanjeev.khanna@nyu.edu}.} \and Aaron Putterman\thanks{School of Engineering and Applied Sciences, Harvard University, Cambridge, Massachusetts, USA. Supported in part by a Jane Street Graduate Research Fellowship, the Simons Investigator Awards of Madhu Sudan and Salil Vadhan and AFOSR award FA9550-25-1-0112. Email: \texttt{louiealbp@gmail.com}.} \and Junkai Song\thanks{Courant Institute, New York University, New York, NY, USA. Supported in part by NSF award CCF-2625203. Email: \texttt{junkaisong@nyu.edu}.}}

\begin{document}
\maketitle

\begin{abstract}
    We show that there is a deterministic linear time algorithm for the minimum spanning tree problem, thereby settling this problem, and providing the first improvement since the work of Chazelle (JACM 2000). Our work proceeds by showing that the deterministic comparison complexity of the minimum spanning tree problem is linear, whereby a known reduction of Pettie and Ramachandran (JACM 2002) immediately implies a linear time algorithm.
\end{abstract}

\pagenumbering{gobble}

\clearpage

\pagebreak

\tableofcontents

\pagebreak

\pagenumbering{arabic}

\section{Introduction}

The \emph{minimum spanning tree} (MST) problem is a cornerstone of theoretical computer science and graph algorithms. 
Given a connected undirected graph $G=(V,E)$ with $n$ vertices, $m$ edges, and a real-valued weight function $w\colon E\to\mathbb{R}$, the MST problem asks for a spanning tree of~$G$ with minimum total weight.\footnote{In general, for a disconnected graph, one asks for a \emph{minimum spanning forest} (MSF), which simply contains an MST of each connected component.}

The MST problem has a long and storied history, with the first work studying it often being attributed to Bor\r{u}vka \cite{boruuvka1926jistem,BoruvkaElectric1926}, who encountered the problem during an effort to design an efficient scheme for the electrification of Moravia.\footnote{Presumably, the vertices were villages, and edge weights represented the cost of connecting the power grid between two towns.} This study led to the development of the  first efficient algorithm for the MST problem \cite{boruuvka1926jistem,BoruvkaElectric1926}, and to this day it remains one of the most extensively studied problems in the area of combinatorial algorithms. Several textbook algorithms based on natural greedy rules, including Bor\r{u}vka's algorithm, admit near-linear $O(m\log n)$-time implementations. However, despite its rich history, the most basic question about the MST problem has remained unresolved:

\begin{center}
\emph{Does the MST problem admit a
deterministic algorithm \\with $O(m)$ worst-case running time?}
\end{center}

Naturally, the above question has seen significant attention, and as a consequence, several weaker linear-time bounds were already known. Klein and Tarjan~\cite{KleinTarjan1994}, and later Karger, Klein, and Tarjan~\cite{KargerKleinTarjan1995},  gave a \emph{randomized} algorithm that always returns an MST and runs in linear time both in expectation and with high probability. Fredman and Willard~\cite{FredmanWillard1994} gave a deterministic linear-time algorithm for MST with word-sized integer weights on a word RAM. Similarly, deterministic linear-time \emph{verification} of a supplied spanning tree was known by the early 1990s~\cite{DixonRauchTarjan1992}.

We study the most general version of the problem, where the algorithm may only compare edge weights in pairs, with each comparison taking constant time, and otherwise treats the weights as opaque keys (thus, the algorithm works for even the most unstructured weighted settings). In this setting, the best previously known explicit worst-case bound for deterministic comparison-based construction was $O(m\alpha(m,n))$, due to Chazelle~\cite{chazelle2000minimum}. In this context, $\alpha$ is a parameterization of the extremely slowly-growing \emph{inverse-Ackermann} function. In a later work, Pettie and Ramachandran~\cite{pettie2002optimal} established that optimal deterministic running time is governed by $\mathcal{T}^*(m,n)$, which we use to denote the worst-case deterministic \emph{comparison complexity} of minimum spanning forests on simple graphs with $m$ edges and $n$ vertices (see \cref{sec:history}). Consequently, a deterministic linear-time comparison-based MST algorithm exists if and only if $\mathcal{T}^*(m,n)=O(m)$. However, this work of \cite{pettie2002optimal} left open a tight bound on $\mathcal{T}^*(m,n)$.

In this paper, we settle the above question by proving the following theorem.

\begin{theorem}\label{thm:main}
There is a uniform deterministic algorithm that, on every connected undirected multigraph with $n$ vertices and $m$ edges with real weights, finds an MST in $O(m)$ worst-case time, accessing edge weights only through pairwise comparisons.
\end{theorem}

\subsection{A brief history of the MST problem} \label{sec:history}

We now give a brief overview of prior work on the MST problem, starting with the seminal work of \cite{boruuvka1926jistem,BoruvkaElectric1926}. Bor\r{u}vka's 1926 algorithm works in phases: starting with a graph $G = (V, E)$, each vertex selects its incident edge with the smallest weight. The algorithm then \emph{commits} to keeping these edges in the MST, and effectively \emph{contracts} on these committed edges, treating the created components as their own vertices. In this way, every phase at least halves the number of vertices, so after $O(\log n)$ phases, each using $O(m)$ comparisons, one obtains an MST. In 1930, Jarn\'ik~\cite{Jarnik1930} described a tree-growing alternative. Indeed, this work proposed starting from a single vertex and repeatedly adding the cheapest edge leaving the current tree. This algorithm was later rediscovered by Prim~\cite{Prim1957} and by Dijkstra~\cite{Dijkstra1959} and is now commonly called Prim's algorithm. Kruskal~\cite{Kruskal1956} developed yet another alternative. In his work, he proposes scanning the edges in \emph{increasing order of weight} and keeping each edge that closes no cycle. Naturally, Kruskal's algorithm relies on sorting the underlying edges (in addition to maintaining a connectivity data structure) and so also requires an $O(\log n)$ overhead. For a more complete discussion of these early algorithms and their rediscoveries, we encourage the reader to the work of Graham and Hell~\cite{GrahamHell1985} trace these developments. Similarly, Ne\v{s}et\v{r}il, Milkov\'a, and Ne\v{s}et\v{r}ilov\'a~\cite{NesetrilMilkovaNesetrilova2001} give English translations of both of Bor\r{u}vka's 1926 papers.

\begin{table}[htbp]
\centering
\small
\renewcommand{\arraystretch}{1.12}
\begin{tabular}{@{}l@{\quad}p{.405\linewidth}@{\quad}p{.37\linewidth}@{}}
\toprule
\textbf{Year} & \textbf{Algorithm} & \textbf{Running-time guarantee} \\
\midrule
1926 & Bor\r{u}vka~\cite{boruuvka1926jistem,BoruvkaElectric1926} & $O(m\log n)$ \\
1930 / 1957 & Jarn\'ik~\cite{Jarnik1930} / Prim~\cite{Prim1957} & $O(m+n\log n)$ with Fibonacci heaps \\
1956 & Kruskal~\cite{Kruskal1956} & $O(m\log n)$ \\
1975 & Yao~\cite{Yao1975} & $O(m\log\log n)$ \\
1975 & Johnson~\cite{Johnson1975} & $O(m\log_d n)$ \\
1976 & Cheriton and Tarjan~\cite{CheritonTarjan1976} & $O(m\log(\log_d n))$ \\
1984 / 1987 & Fredman and Tarjan~\cite{FredmanTarjan1987} & $O(m\beta(m,n))$ \\
1984 / 1986 & Gabow, Galil, and Spencer~\cite{GabowGalilSpencer1984}, later with Tarjan~\cite{GabowEtAl1986} & $O(m\log\beta(m,n))$ \\
1997 & Chazelle~\cite{Chazelle1997} & $O(m\alpha(m,n)\log\alpha(m,n))$ \\
2000 & Chazelle~\cite{chazelle2000minimum} & $O(m\alpha(m,n))$ \\
2000 / 2002 & Pettie and Ramachandran~\cite{pettie2002optimal} & $O(T^*(m,n))$, optimal \\
\midrule
\multicolumn{3}{@{}l}{\emph{Other settings}} \\
1990 / 1994 & Fredman and Willard~\cite{FredmanWillard1994} & $O(m)$ for word-sized integer weights on a word RAM \\
1994 / 1995 & Klein and Tarjan~\cite{KleinTarjan1994}, then Karger, Klein, and Tarjan~\cite{KargerKleinTarjan1995} & $O(m)$ in expectation and with high probability \\
\bottomrule
\end{tabular}
\caption{Selected MST algorithms for connected simple graphs where the first block gives deterministic comparison-based worst-case bounds. In the above, we use $d:=2+m/n$, and $\beta(m,n) := \min\{\,i\ge 1 : \log^{(i)} n \le m/n\,\}$.}
\label{tab:history}
\end{table}

All of the above-mentioned algorithms rest on two dual principles. Assume, as we may by breaking ties consistently, that the edge weights are \emph{distinct}, so that the MST is \emph{unique}. The \emph{cut property} states that the lightest edge crossing any cut belongs to the MST; the \emph{cycle property} states that the heaviest edge on any cycle does not.

Tarjan~\cite{Tarjan1983} cast these two conditions above as coloring rules. On a cut crossed by no blue edge, color the lightest uncolored crossing edge blue. On a cycle containing no red edge, color the heaviest uncolored edge red. Performing a any sequence of moves in keeping with the above conditions, once every edge is colored, the blue edges necessarily form the MST.

\paragraph{Beating sorting.}
Since the MST depends only on the \emph{relative order} of the weights, the natural cost measure of finding the MST is the number of comparisons (together with the amount of work needed to decide which comparisons to make). On simple graphs, each of the aforementioned classical algorithms runs in $O(m\log n)$ time. Kruskal's algorithm sorts the edges in $O(m\log m)=O(m\log n)$ time, Prim's algorithm performs $O(m)$ operations on a binary heap of at most $n$ vertices, and Bor\r{u}vka's algorithm may need $\Theta(\log n)$ linear-time phases.

Progress towards linear time MST algorithms has largely happened by first focusing on \emph{dense} instances. Indeed, on dense simple graphs with $m=\Theta(n^2)$, an $O(n^2)$ array implementation of Prim's algorithm already runs in linear time. This was further improved in the mid-1970s, which brought about algorithms that improved on the $O(m\log n)$ comparison bound even for \emph{sparse graphs}. For instance, Yao~\cite{Yao1975} combined Bor\r{u}vka's phases with the then-new linear-time selection algorithm~\cite{BFP+72}, partially sorting the adjacency lists to obtain $O(m\log\log n)$ time. Later work  by Cheriton and Tarjan~\cite{CheritonTarjan1976} obtained the sharper bound $O(m\log(\log_d n))$ (where $d=2+m/n$), which is even \emph{linear} when $m=\Omega(n^{1+\varepsilon})$ for any constant $\varepsilon>0$, together with a linear-time algorithm for planar graphs.

In the 1980's, the seminal work of Fredman and Tarjan~\cite{FredmanTarjan1987} introduced \emph{Fibonacci heaps}. Using these, Prim's algorithm can be implemented to run in $O(m+n\log n)$ time, which is linear for any $m=\Omega(n\log n)$. In fact, using these heaps, Fredman and Tarjan were able to produce an even faster \emph{phased} algorithm. In each phase of this new algorithm, trees are grown one at a time in a manner similar to Prim's algorithm. Each tree starts from a vertex of the current contracted graph that is not yet in any tree and storing its neighboring vertices in a heap keyed by their cheapest connecting edges. A tree stops growing when its heap exceeds a threshold, when it reaches a tree grown earlier in the phase, or when its heap becomes empty. The algorithm then contracts the resulting trees and repeats this procedure with a larger threshold. By carefully optimizing these thresholds, this yields a deterministic $O(m\beta(m,n))$ time algorithm, where
\[
  \beta(m,n) = \min\{\,i\ge 1 : \log^{(i)} n \le m/n\,\},
\]
(note that $\log^{(i)}$ denotes $i$ iterations of the base-two logarithm). In particular, this means that $\beta(m,n)\le\log^* n$, and one can even show that the algorithm runs in \emph{linear time} whenever $m\ge n\log^{(k)} n$ for a fixed positive integer~$k$. Gabow, Galil, and Spencer~\cite{GabowGalilSpencer1984} later refined this approach in 1984 to obtain a deterministic $O(m\log\beta(m,n))$ time algorithm.

\paragraph{Non-greedy algorithms and the $O(m\alpha(m,n))$ bound.} Progress beyond this $O(m\log\beta(m,n))$ running time required a fundamentally new idea. This was ultimately supplied in 1997, when Chazelle \cite{Chazelle1997} gave a breakthrough deterministic algorithm running in $O(m\alpha(m,n)\log\alpha(m,n))$ time. Here $\alpha$ is a two-parameter inverse-Ackermann function as in Tarjan's analysis of union--find~\cite{Tarjan1975}, and is in general, an extremely slowly growing function. Chazelle subsequently improved the bound to $O(m\alpha(m,n))$ in \cite{chazelle2000minimum}, along with an independent, similar bound developed by Pettie \cite{Pettie1999} (still building on \cite{Chazelle1997}).

Chazelle's algorithm crucially \emph{departs} from the greedy paradigm. Its main tool is the so-called \emph{soft heap}~\cite{ChazelleSoftHeap2000}, a meldable priority queue that can \emph{corrupt} a key by increasing its value. For an error parameter $0<\varepsilon\le 1/2$, after $N$ insertions, the guarantee is simply that \emph{at most} $\varepsilon N$ items currently in the heap have \emph{corrupted} keys. Every operation takes constant amortized time except insertion, which takes $O(\log(1/\varepsilon))$ amortized time. With these soft heaps in hand, the algorithm first computes a spanning tree and then refines it until it becomes an MST. The ultimate algorithm is comparison-based, runs on a pointer machine, and makes no numerical assumptions about the weights, and, until now, remained the best explicit bound on the deterministic complexity of finding an MST.

\paragraph{Verification algorithms.}
In parallel to these works designing better MST algorithms, another line of work has concerned \emph{verification}:  i.e., the task of deciding whether a given spanning tree~$T$ is minimum. By the aforementioned cycle property, this effectively amounts to checking that every non-tree edge is heavier than every edge on the $T$-path between its endpoints. In his work \cite{Tarjan1979}, Tarjan showed how to solve this path-maximum problem in $O(m\alpha(m,n))$ time using path compression on balanced trees. Koml\'os~\cite{komlos1985linear} then refined this, and showed that a \emph{linear} number of comparisons suffices for verification. However, his method did not give a \emph{linear-time} way to decide which comparisons to make. This was then first achieved by Dixon, Rauch, and Tarjan \cite{DixonRauchTarjan1992}, who obtained a linear-time implementation on a unit-cost RAM by combining Koml\'os's algorithm with table lookup for small subproblems and with Tarjan's earlier method. King \cite{king1997simpler} later gave a simpler implementation of this algorithm, and ultimately,  Buchsbaum et al.~\cite{BKRW1998,BGKRTW2008} even achieved linear time on a pointer machine.

Verification also turned out to be useful for \emph{constructing} MSTs. For any forest $F\subseteq G$, let us call an edge \emph{$F$-heavy} if it is heavier than every edge on the $F$-path between its endpoints, and \emph{$F$-light} otherwise. By the cycle property, $F$-heavy edges can be discarded, and an adaptation of the verification algorithm identifies all of them in linear time~\cite{DixonRauchTarjan1992,KargerKleinTarjan1995}. This raised the hope that this \emph{filtering} could be a stepping stone for faster deterministic algorithm. Pettie~\cite{Pettie2006} ruled out one natural approach by proving that \emph{online} verification, in which path-maximum queries arrive one at a time, indeed requires $\Omega(m\alpha(m,n)+n)$ comparisons. 

\paragraph{Randomized algorithms.}
Despite this barrier proved by \cite{Pettie2006}, it is nevertheless the case that combining this \emph{filtering} step with random sampling is the key idea needed to get an $O(m)$ time \emph{randomized} MST algorithm. Karger \cite{Karger1993} initiated this approach, whereby he combined random sampling with verification to obtain an MST algorithm running in $O(m+n\log n)$ time with high probability. Klein and Tarjan~\cite{KleinTarjan1994} refined this, obtaining an expected linear time algorithm in 1993.  A subsequent work of Karger, Klein, and Tarjan~\cite{KargerKleinTarjan1995} provides the same result but with an \emph{even stronger} probability bound. Their algorithm alternates between so-called ``Bor\r{u}vka phases'', which shrink the number of vertices, with the new idea of sampling and filtering, which shrink the number of edges. The crucial lemma which underlies their result shows that if each edge is sampled  independently with probability $0<p\leq 1$ and $F$ is the sample's minimum spanning forest, then the expected number of $F$-light edges of~$G$ is at most $n/p$. This idea of using Bor\r{u}vka phases to reduce vertices, and ``filtering'' to reduce edges, will also re-appear in the algorithm we present below (though the filtering will naturally require an entirely different approach to achieve a deterministic guarantee).

\paragraph{Stronger models and structured inputs.}
Earlier deterministic algorithms also achieved linear time by exploiting the representation of edge weights or the structure of the input graph. Fredman and Willard's algorithm~\cite{FredmanWillard1994} runs in linear time on a unit-cost word RAM when each integer weight fits in a machine word. Its atomic heaps exploit word-level operations on the keys. Within the comparison model, deterministic linear-time algorithms were known for planar graphs~\cite{CheritonTarjan1976}, for each fixed proper minor-closed class of simple graphs~\cite{Mares2004}, and for sufficiently dense graphs~\cite{FredmanTarjan1987}, as we discussed above.

\paragraph{An optimal algorithm and its comparison complexity.} Pettie and Ramachandran~\cite{pettie2002optimal} made the first explicit link between the number of comparisons required to find an MST and running time. To make this formal, let us fix a simple undirected graph $H$. A \emph{comparison tree} for $H$ is a decision tree where each internal node compares two edge weights and whose leaves are labeled with spanning forests of $H$. We say that it is correct if, for every possible assignment of distinct weights to $H$, the leaf reached is labeled with the resulting MSF. The depth of the comparison tree is then exactly the worst-case number of comparisons required to find an MSF of $H$. With this, we can define
\[
  \mathcal{T}^*(m,n) \;=\; \max_H\, \min_{T}\, \operatorname{depth}(T),
\]
where $H$ ranges over all simple undirected graphs with $m$ edges and $n$ vertices, including disconnected ones, and $T$ ranges over all correct comparison trees for $H$.

Every deterministic comparison-based MSF algorithm induces a correct comparison tree for each fixed topology $H$, so on graphs with $m$ edges and $n$ vertices it makes at least $\mathcal{T}^*(m,n)$ comparisons in the worst case. Pettie and Ramachandran gave a uniform pointer-machine algorithm that matches this bound. Indeed, on connected inputs, their algorithm runs in $O(\mathcal{T}^*(m,n))$ time\footnote{This is why the definition of $\mathcal{T}^*$ allows disconnected graphs. A connected graph with $m=n-1$ is its own MST, so over connected graphs alone we would have $\mathcal{T}^*(n-1,n)=0$, whereas reading such an input already takes $\Omega(n)$ time.} and it constructs the \emph{optimal} comparison trees that it uses, relying on a careful use of soft heaps, Bor\r{u}vka steps, and brute force computation of optimal comparison trees on small sub-instances.

For $m,n>2$, the best previously known bounds were
\[
  \mathcal{T}^*(m,n)=\Omega(m) \qquad\text{and}\qquad \mathcal{T}^*(m,n)=O(m\alpha(m,n)),
\]
leaving open whether a linear number of comparisons always suffices. 

In this work, we show that it does: $\mathcal{T}^*(m,n)=O(m)$. Together with the algorithm of Pettie and Ramachandran~\cite{pettie2002optimal}, this yields \cref{thm:main}.

\subsection{Technical Overview}

\paragraph{Why Edge Filtering is Key}

As alluded to above, the key new technical ingredient that we introduce is a mechanism for \emph{efficiently finding} many ``redundant edges'' (i.e., edges which are the heaviest edges in a cycle in the graph). To properly describe how the algorithm works, we first introduce the notion of being \emph{supported}. Indeed, for a graph $G = (V, E)$ and a forest $F \subseteq E$, we say that an edge $e \in E$ is \emph{supported} by $F$, if the endpoints of $e$ are connected by $F_{\leq w(e)}$. In words, this is simply asking that if one adds the edge $e$ to the forest $F$, then the resulting cycle uses edges of weight only $\leq w(e)$. With this, we can present the formal statement of our key subroutine, namely, the certificate procedure:

\begin{lemma}[Certificate Procedure]
Let $G$ be an undirected multigraph with $n$ vertices, $m$ edges, and no isolated vertices. For every fixed constant $\varepsilon>0$, there exists a deterministic certificate procedure that outputs a forest $F\subseteq E(G)$. The procedure may invoke exact MSF calls on edge sets $E_1,\ldots,E_\ell$, each defined on a subset of $V(G)$. Apart from these calls, the procedure performs $O_\varepsilon(m)$ weight comparisons. Let $U(F)\subseteq E(G)$ denote the set of edges unsupported by $F$. The procedure guarantees that
\[
|F|+|U(F)|+\sum_{i=1}^{\ell}|E_i| \leq \varepsilon m+C_\varepsilon\sqrt{nm},
\]
where $C_\varepsilon$ is a constant depending only on $\varepsilon$.
\end{lemma}

Let us now unpack this lemma: we fix $\varepsilon$ to be a small constant (in the final proof, we may even choose it to be $1/8$). The lemma then guarantees that there is a procedure which produces a forest $F$ which in fact supports \emph{many} of the original edges in the graph. Because these original edges are supported, this then provides a proof that (unless they themselves are in the forest), they are \emph{not} present in the MST. Viewing this as a filtering procedure, this means that we can reduce the number of edges in the graph from $|E(G)|$ all the way down to $|F| + |U(F)|$.

This now explains why the left side of the expression in the above lemma is meaningful. Ultimately, we will recurse on a graph with $|F| + |U(F)|$ edges, which is exactly the number of edges left over after filtering. The second piece of the expression on the left is the actual \emph{MST recursive calls} that are used in the computation of the certificate. Thus, the combined size of all recursive calls that are performed is bounded by $|F|+|U(F)|+\sum_{i=1}^{\ell}|E_i| \leq \varepsilon m+C_\varepsilon\sqrt{nm}$.

This expression itself is still not perfect however. On the right hand side, the $\varepsilon \cdot m$ term is quite small, but $C_\varepsilon\sqrt{nm}$ term can be quite large, especially when $n$ approaches $m$. This is the key place where Bor\r{u}vka phases come into play. A single Bor\r{u}vka phase uses $O(m)$ comparisons and \emph{halves} the number of vertices in the graph. Because $C_{\varepsilon}$ is just a constant, we can then apply a \emph{large enough constant} number of Bor\r{u}vka phases to reduce $n$, and thereby ensure that $C_\varepsilon\sqrt{nm} \leq m/4$. All together, this will then guarantee that the combined size of all recursive calls satisfies $|F|+|U(F)|+\sum_{i=1}^{\ell}|E_i| \leq m/2$.

In order to finish off the argument, we still need to bound the total number of comparisons that are made. This just requires summing up the number of comparisons across all pieces of the algorithm: (1) the certificate procedure itself makes at most $O_{\varepsilon}(m)$ weight comparisons, (2) performing a constant number of Bor\r{u}vka phases requires only $O(m)$ weight comparisons, and finally, (3) by using the work of \cite{komlos1985linear}, it turns out that \emph{identifying} the set of unsupported edges (given a forest $F$) can be done in only $O(m)$ comparisons. 

All together, this leads to a recursive relationship of the form $T(m) = T(m/2) + O(m)$, which gives a bound of the form $T(m) = O(m)$, as we desire.

\paragraph{Intuition for the Certificate Procedure} With the above in place, the remaining ingredient (and indeed the key roadblock) is to understand \emph{how} we can actually build (and prove the correctness of) the certificate procedure. At a very high level, the procedure operates by repeatedly \emph{projecting} vertices onto smaller and smaller vertex sets. In each such projection, the goal is for a vertex $v$  to be assigned to a vertex $\pi(v)$, such that the weight of the edge connecting $v$ and $\pi(v)$ is small. In this way, for an edge $(u,v)$, if $u$ and $v$ are repeatedly projected, eventually reaching a small enough vertex set where one can find a low-weight path between their respective projections, then this should in fact produce a \emph{cycle} involving $(u,v)$, where $(u,v)$ is the highest-weight edge. 

In a bit more detail, this is achieved by maintaining a subset $P$ of the vertices, called \emph{portals}. For each vertex $v$, the goal is to maintain an incident edge $\sigma(v)$ connecting $v$ to a portal, called the \emph{spoke} of $v$. Notationally, we let $\pi(v)$ denote the portal at the other endpoint of the spoke. It is also possible that $v$ may not have a spoke, in this case we write $\sigma(v)=\pi(v)=\,\perp$. 
Now, for an edge $e=(u,v)$ whose endpoints both have spokes, the algorithm may \emph{project} $e$ to its portals $\pi(u)$ and $\pi(v)$. In the process, the algorithm represents the edge $(u,v)$ with an edge of the same weight between $\pi(u)$ and $\pi(v)$. 

Crucially, ensuring that $(u,v)$ is the largest weight edge in the cycle \emph{requires} that $(u,v)$ is heavier than both of its spokes. For this reason, we let $p(e)=\max\{w(\sigma(u)),w(\sigma(v))\}$ be the larger of the two spoke weights, and we say the projection is \emph{valid} if $p(e)\le w(e)$. Because the algorithm repeatedly projects edges, the certificate procedure itself will be a \emph{recursive} algorithm. At each level of the recursion, the input to the certificate procedure is an undirected multigraph $G$. Each edge $e\in E(G)$ has a weight $w(e)$ and a \emph{projection record}, which stores, for each preceding recursion level, the pair of spokes used to project $e$ at that level. We say that $e$ is \emph{admissible} if every spoke in its projection record has weight at most $w(e)$, or equivalently, if all projections of $e$ at preceding recursion levels are valid. 

With this intuition established, there are still several questions we must answer like (1) how do we decide where to project vertices? (2) how do we preserve bounds on the number of inadmissible edges? (3) how do we ensure that we do not use too many comparisons?

\paragraph{The Certification Algorithm} Unfortunately, there is no single, modular description where each of the above questions is answered separately. Instead, we must introduce the entire certification algorithm, and then separately reason about its (frequently interfering) properties. We now describe this algorithm, and encourage the reader to separately view the formal pseudocode of the algorithm in \cref{alg:certificate} after understanding the essence of its behavior.

The algorithm takes as input a graph $G$ with $n$ vertices and $m$ edges, and initializes its forest set $F = \varnothing$, its portal set $P = \varnothing$, and its set $R = \varnothing$ of rejected (inadmissible edges). Similarly, every vertex is given a starting empty spoke $\sigma(v) =\, \perp$. The algorithm is then broken into $t^2$ \emph{disjoint rounds}, which are then organized into $t$ blocks of $t$ rounds ($t$ itself is a parameter we will choose later, which relates to the recursion depth).
The intuition is that the edges $E(G)$ will now be partitioned among the $t^2$ rounds, with each edge being assigned to a single round (we call this the \emph{schedule} of the algorithm). This schedule will designate \emph{when the edge} $e$ is permitted to act -- a priori it may seem excessive to keep track of the algorithm on such a fine-grained level, but designing a linear comparison algorithm unfortunately requires extensive care of exactly when and where edges are permitted to have their weights queried. Importantly, we cannot afford to have many edges re-appearing in different recursive calls, thereby increasing the comparison complexity. 

To this end, we now further restrict each edge. In the round where it is selected to act, it has only two possible actions that it can take:
\begin{enumerate}
    \item The $\textsf{find}$ action searches for a valid projection.
\item  The $\textsf{project}$ action projects $e$ to $(\pi(u),\pi(v))$ using the current spokes, \emph{regardless} of whether the projection is valid. Crucially however, those edges taking the $\textsf{project}$ action will enter a child recursive instance (unlike the edges taking the $\textsf{find}$ action).
\end{enumerate}
It is worth noting that the choice between doing a $\textsf{find}$ action vs. a $\textsf{project}$ action is \emph{completely deterministic}. Whenever an edge selects the $\textsf{find}$ action, it is added to the current \emph{batch} $B$ of edges. Otherwise, if an edge selects the $\textsf{project}$ action, we project the edge along its spokes, and \emph{recursively} invoke the certificate algorithm. 

Finally, at the end of every block of $t$ rounds, the cumulative \emph{batch} of edges $B$ that has been collected so far is processed. This produces a set $P'$ of new portals which is combined with $P$, a set $D$ of so-called ``declared'' edges which are used to update the forest $F$, along with some potentially rejected edges $R'$, which are added to the rejected set.

With this context established, we can now more concretely ask the above questions, which we present below. The remainder of the technical overview then describes the intuition for how each of these questions is answered. Note that several other details (and natural questions) arise in the design of this algorithm, but for the sake of brevity and clarity in this overview, we postpone their discussion to the actual algorithm analysis (see \cref{sec:ceritificateAlgorithm}):

\begin{enumerate}
    \item How do we decide which action an edge should take?
    \item Why does the final forest produced by this procedure not have too many unsupported edges?
    \item How do we select the schedule for the algorithm? Recall that we are not permitted to use randomness here!
\end{enumerate}

\paragraph{Deciding Edge Actions}

We now must first discuss \emph{which of the two actions} an edge should take when it is scheduled to act in a given round. This is captured by our ``preferred action'' framework, which is essentially just a deterministic set of rules which governs what action is preferred by an edge in a given round. 

Recall that at the start of each round, every vertex is already assigned a spoke (or potentially $\perp$), and additionally, that there is a candidate forest $F$ that the algorithm has computed so far.  We then automatically say that the \emph{preferred action} of an edge $e=(u,v)$ is $\textsf{find}$ if either endpoint has no spoke or if both spokes lead to the same portal (if they are the same portal, then this is intuitively already making a cycle). Otherwise, when the spokes lead to distinct portals, we have a more careful tradeoff. We designate the preferred action to be $\textsf{find}$ if $F_{\leq p(e)}$ connects $u$ and $v$, and $\textsf{project}$ otherwise (recall that here $p(e)$ is the weight of the heavier spoke). This already determines the rule for how an edge should ideally act.

We note that there is already one important subtlety which arises: just because it is \emph{theoretically possible} to determine an edge's preferred action, this does not imply that there is an efficient comparison algorithm for computing it! This ends up becoming a material challenge in the design of the algorithm, and thus we must ultimately resort to \emph{estimating} the preferred action of each edge.  The estimate will end up being always correct when  either endpoint has no spoke or if both spokes lead to the same portal (as this does not require comparisons), and otherwise is \emph{mostly correct} at deciding between $\textsf{find}$ and $\textsf{project}$. We omit an extensive discussion of this estimation procedure here, but it relies on a careful bound on the \emph{number of possible preference vectors} (i.e., a complete enumeration of all possible joint preferences that can appear in a given round), along with a choice of a universal set of edges which can be used to distinguish two preference vectors that are far apart. Roughly speaking, this reduces the task of figuring out the preferences of \emph{all edges} to figuring out the preferred actions for a \emph{smaller set} of edges, while still mostly retaining correctness on the other edges. 

\paragraph{Bounding Unsupported Edges}

For now, let us ignore this preceding detail about estimating the true preferred action of each edge in each round, and instead assume we have access to them. In this ideal situation, we must still show that \emph{choosing preferred actions} actually leads to a good outcome in the $\Cert$ algorithm. 

Towards establishing this, we must first introduce the notion of good and bad rounds: 
\begin{definition}
A round is \emph{bad} for $e$ if either of the following holds:
\begin{itemize}
    \item It is the last round in its block with a \textsf{project} preference.
    \item The preferred action is \textsf{find}, and this block is the \emph{last block} containing any \textsf{find} preference for this edge. 
\end{itemize}
\end{definition}

In more detail, within a single block, the preferences of an edge always look like PPPPFFF where P corresponds to \textsf{project} and F to \textsf{find} (in the sense that they are monotonically changing from P's to F's). The final P in any such block is said to be bad. Between blocks, preferences can transition back to Ps -- i.e., something like 
\[
\dots \dots \quad \text{PPPPFFF PPFFFFF PPPPPPP }
\]
is possible. In this example, all the F's from the second to last block are bad. The crucial fact we end up showing is the following: 

\begin{lemma}
Consider an instance $\Cert(G)$ and an edge $e\in E(G)$. Suppose that, for an edge $e$, the edge $e$ and all of its projections throughout the recursion are \emph{never scheduled to act in a bad round}. Then, $e$ is supported by the final forest $F$ returned by the algorithm.
\end{lemma}

In other words, whenever an edge is rejected or otherwise not supported by the returned forest $F$ from a call to $\Cert(G)$, the only way that this is possible is that the edge or one of its projections was scheduled to act in a round where its preferred action was bad. Fortunately for us, we also now have a concrete understanding of how many bad rounds there are for a single edge. Out of the $t^2$ total rounds, there are only $t$ bad choices of a bad round for preferred action P (one in each block) and only $t$ bad choices for preferred action F (all rounds in the final block), leading to a total of $2t$ bad rounds. This only corresponds to a $2/t$ fraction of rounds being bad, which is sufficient for our final bound that there are not too many unsupported edges. 

\paragraph{Finding a Good Schedule}
Nevertheless, the preceding discussion raises a separate concern. Even though in \emph{expectation}, the number of bad rounds for each edge is small, how do we find a deterministic choice of schedule which \emph{realizes} this behavior? This question is really at the heart of designing our linear time MST algorithm. The main difficulty is a subtle circular dependence: the preferred actions of each edge \emph{depends} on the schedule, which in turn determines which rounds are good or bad. To reason about this dependence, we introduce an object called the \emph{history}.

\begin{definition}[History]
Consider a call $\Cert(G)$ on a graph $G$ with $n$ vertices and $m$ edges. We index the $t^2$ rounds by $i\in [t^2]$. We define the history to be
\[
H=(\sigma_i,F_i,a_i)_{i\in [t^2]},
\]
where:
\begin{itemize}
    \item $\sigma_i$ is a vector of length $n$, where $\sigma_i(v)\in E(G)\cup{\perp}$ specifies the spoke of vertex $v$ at the start of round $i$.
    \item $F_i\subseteq E(G)$ is the forest $F$ maintained by the algorithm at the start of round $i$.
    \item $a_i$ is a vector of length $m$, where $a_i(e)\in \{\textsf{find},\textsf{project}\}$ specifies the \emph{true preferred action} of edge $e$ at the start of round $i$.
\end{itemize}
\end{definition}

Note that the history completely determines which rounds are good or bad. Our analysis will begin by showing that the number of histories is \emph{not too large}. This then enables us to show that there exists a \emph{single deterministic schedule} that ensures not too many edges are assigned to bad rounds. 

The bound on the number of histories, though not complicated, is subtle. The number of choices for $\sigma_i$ and $F_i$ is easy to bound. Each vertex in $\sigma_i$ has $\mathrm{deg}(v)+1$ choices for its spoke, and each forest $F_i$ has $\leq \binom{m+n-1}{n-1}$ choices. Taking the product over all choices of $i \in [t^2]$ and all vertices, this gives a simple bound of $e^{O(t^2 \sqrt{mn})}$ possibilities for $\sigma_i, F_i$. It remains still to bound the number of possible \emph{preference vectors} $a_i$. 

For this bound, let us fix the spokes $\sigma_i$ and the forest $F_i$ at the start of round $i$. For each edge $e=(u,v)\in E(G)$, recall that its preferred action is $\textsf{find}$ if (1) either endpoint has no spoke, (2) $\pi(u)=\pi(v)$, or (3) $F_{\le p(e)}$ connects $u,v$, otherwise, it is $\textsf{project}$. The first two cases are completely determined by the spokes (and so do not contribute to a larger number of histories), so it remains to analyze case (3).

Consider a ground set $T$ of size at most $2n$ consisting of all the spokes and edges of $F_i$. The total order is completely determined by edge weights. For each edge $e=(u,v)\in E(G)$, let $S_e\subseteq T$ consist of the spokes $\sigma_i(u),\sigma_i(v)$ and all edges on the $u$ to $v$ path in $F_i$, if such a path exists. The preferred action of $e$ is \textsf{find} precisely when the maximum element of the set $S_e$ is a spoke. Indeed, this occurs precisely when the heaviest spoke is heavier than every edge on the $u$ to $v$ path in $F_i$, which is equivalent to $u,v$ being connected in $F_{\leq p(e)}$. Conversely, if the maximum weight edge in $S_e$ is a path edge, then $u$ and $v$ are not connected in $F_{\leq p(e)}$. Note that here we are use the fact that there is a \emph{unique} path between $u$ and $v$ in $F$ (if one exists). 

The takeaway from the above is that the preferred actions are entirely determined by the \emph{maxima} of the sets $S_e$. A known result (first appearing due to Fredman in \cite{graham1980information}) then states that in a universe of size $N$, and $M$ subsets of $[N]$, the number of different ``maximizing patterns'' (and thus, the number of different preference vectors) is at most $e^{O(\sqrt{NM})}$. Taking the bound across all $t^2$ rounds then bounds the number of histories by $e^{O(t^2\sqrt{mn})}$. 

This is the first place where the parameter guarantees from the certificate procedure come into focus, with this $\sqrt{mn}$ being the same factor that appears. Indeed, with this bound on the number of histories, a careful probabilistic argument shows that we can \emph{deterministically} (without any comparison queries) decide on a schedule which will not schedule more than $O(m/t + t^2\sqrt{mn})$ many edges in bad rounds.

This marks one of the crucial insights in the paper, and indeed the main bound which allows us to reach linear time. We do remark that there are several hurdles which we have overlooked (for instance, bounding the recursive complexity of calling the \Cert\ procedure on portal graphs, or bounding the number of queries required to determine which edges are admissible and not, and careful accounting of where rejected edges are sent). We encourage readers to visit \cref{sec:ceritificateAlgorithm} to see the complete algorithm pieced together.

\subsection{Acknowledgments}

We thank Vijaya Ramachandran for drawing our attention to her joint work with Seth Pettie~\cite{pettie2002optimal} during the first author's visit to the University of Texas at Austin in spring 2026. It was
then that we began working on the problem of designing a deterministic linear-time MST algorithm.

\subsection{AI Acknowledgments}

A. P. would like to thank OpenAI for providing the credits to Astra 6.2 which were used in generating this result.

Indeed, the main ideas in this paper were entirely generated by AI -- specifically 6.2 Astra, on a mixture of Ultra and Max effort -- with minimal human intellectual intervention (see below). The authors then worked to first carefully verify, and then to simplify and clarify the treatment, resulting in the present manuscript. 

The authors had originally tried to query Astra 6 and 6.2 with a direct prompt of achieving deterministic linear comparison complexity for this problem, with no success. The authors then re-tried with the simpler task of proving any bound of the form $o(m \alpha)$, which would be an improvement over the work of Chazelle \cite{chazelle2000minimum}. A team of 15 agents (managed across several Ultra and Max effort settings) succeeded in proving such an improvement after 15 hours, providing a bound of the form $O\left (m \cdot \frac{\alpha}{\log(\alpha)^{1/3}} \right)$. This gave the authors confidence that $O(m)$ is indeed the right answer. Before that, there was some doubt that linear time was the correct answer, largely due to $\Omega(m \alpha)$ lower bounds for related union-find data structures (see \cite{tarjan1979class}), and variants of the MST problem \cite{Pettie2006}. After this, the authors continuously prompted the model to iteratively improve the comparison complexity, resulting in a sequence of small improvements. This took several forms. The first improvement was a bound of the form $O(m \log(\alpha))$, which was subsequently followed by $O(m \log^*(\alpha))$, then $O(m \alpha(\alpha))$, and then $O(m \Gamma(\alpha))$. Here, $\Gamma$ can be seen as the ``inverse of a hyper-Ackermannian function''. See \cite{schmitz2016complexity}, where the function $\Gamma$ is essentially the inverse of $F_{\omega^{\omega}}$. In particular, $\Gamma$ is smaller than \emph{any constant number} of iterated inverse-Ackermann functions. This was then followed by bounds of the form $O(m \log(\Gamma(\alpha)))$, then $O(m \log^*(\Gamma(\alpha)))$, and even the absurdly small $O(m \cdot \Gamma(\Gamma(\alpha)))$. The authors then sent off one final batch of runs which used the entire history of developed algorithms until this point, and asked for a single linear time algorithm, which was finally generated. A collection of these intermediate papers are available at the following URL: \url{https://github.com/Louiealbp/DLMST/tree/main/IntermediateMSTs}.\footnote{There is also a collection of the various linear time algorithms that were ultimately developed at \url{https://github.com/Louiealbp/DLMST/tree/main/OptimalMSTs}.}

Throughout the entire process leading up to the discovery of our main result, human intellectual intervention was limited. The largest such intervention was the advice that the authors did not believe it to be a fruitful direction to pursue ``structural graph theory'' approaches to the problem (whereby the model would study efficient algorithms on graphs which have special structure like low-density, no short cycles, etc.), and repeatedly redirected attempts that got stuck in such pursuits. This guidance was largely due to the fact that the authors themselves had extensively studied structural approaches (in addition to other ideas) for linear time MST algorithms in the spring and summer of this year. 

\section{Preliminaries}
Throughout the paper, $\log$ denotes the natural logarithm. We will use the following standard bound.

\begin{fact} \label{fact:binom}
For every pair of positive integers $n,m$,
\[
\log \binom{m+n}{n} \leq 2\sqrt{mn}.
\]
\end{fact}

\paragraph{Graphs.} Given a graph $G=(V,E)$ and an edge subset $S\subseteq E$, we denote by $G[S]=(V,S)$ the edge-induced subgraph of $G$ with edge set $S$. For each vertex $v\in V$, let $\deg_S(v)$ denote its degree in $G[S]$, and write $\deg_G(v)$ for its degree in $G$.

\paragraph{Minimum spanning forest (\MSF).} For a weighted graph $G$, let $\MSF(G)$ denote its minimum spanning forest. For simplicity, we assume that the graph contains no isolated vertices and that edge weights are distinct with ties broken by edge indices. Under this convention, the MSF is unique.

We recall three standard properties of minimum spanning forests.

\begin{fact}[Cut Property]\label{fact:cutProperty}
    Let $G = (V, E)$ be a weighted graph, and let $S\subseteq V$. If an edge $e$ is the minimum weight edge crossing the cut $(S,V\setminus S)$, then $e\in\MSF(G)$.
\end{fact}

\begin{fact}[Cycle Property]\label{fact:cycleProperty}
    Let $G$ be a weighted graph, and let $C$ be a cycle in $G$. Then the maximum weight edge of $C$ does not belong to $\MSF(G)$.
\end{fact}

\begin{fact}[Contraction property]\label{fact:contractionProperty}
Let $G$ be a weighted graph, and let $S\subseteq\MSF(G)$. Denote by $G/S$ the graph obtained by contracting all edges in $S$. Then, 
\[
    \MSF(G)=S\cup\MSF(G/S).
\]
\end{fact}

For a forest $F=\MSF(G)$ and a weight threshold $k$, let $F_{\leq k}=\{e\in F:w(e)\leq k\}$. We will also use the following property of MSFs: they preserve connectivity at every weight threshold.

\begin{fact} \label{lem:threshold}
Let $G$ be a graph, $S\subseteq E(G)$ and $F=\MSF(G[S])$. For any vertices $u,v$ and threshold $k$, if $G[S]$ contains a path from $u$ to $v$ using only edges of weight at most $k$, then $F_{\leq k}$ connects $u$ and $v$.
\end{fact}

\paragraph{Comparison complexity.}
Our analysis focuses on the number of edge weight comparisons rather than the running time. By the work of Pettie and Ramachandran~\cite{pettie2002optimal}, a deterministic comparison bound for MSF can be translated into a corresponding deterministic running time bound.

\begin{theorem}[Theorem 5.8 of \cite{pettie2002optimal}] \label{fact:PR}
    Let $\mathcal{T}^*(m,n)$ be the deterministic decision-tree complexity of the $\mathrm{MSF}$ problem on graphs with $m$ edges and $n$ vertices. There is a deterministic algorithm that computes the MSF of any graph with $m$ edges and $n$ vertices in $O(\mathcal{T}^*(m,n))$ time.
\end{theorem}

\begin{remark}
As noted in \cite{pettie2002optimal}, their algorithm does not require the full power of the RAM model and can be implemented in the pointer machine model of \cite{tarjan1979class}.
\end{remark}

\paragraph{Bor\r{u}vka steps.} We recall the classical Bor\r{u}vka's algorithm~\cite{boruuvka1926jistem}.

\begin{definition}[Bor\r{u}vka steps]
Given a weighted graph $G=(V,E)$ with $m$ edges, a \emph{Bor\r{u}vka phase} selects, for each vertex $v\in V$, its minimum weight incident edge $e_v$. Let $S=\{e_v:v\in V\}$ be the set of selected edges. The algorithm adds $S$ to the minimum spanning forest and replaces $G$ by the contracted graph $G/S$. 
\end{definition}

By \cref{fact:cutProperty} and \cref{fact:contractionProperty}, we know that $\MSF(G) = S \cup \MSF(G/S)$. Since each minimum weight incident edge $e_v$ can be found using $O(\deg_G(v))$ comparisons, a Bor\r{u}vka step requires $O(m)$ comparisons. Moreover, as every non-isolated vertex is incident to a selected edge, a Bor\r{u}vka step reduces the number of non-isolated vertices by at least a factor of two.

\paragraph{Path maximum queries.} Finally, we will use the following result of Koml\'os~\cite{komlos1985linear} (see also \cite{DixonRauchTarjan1992,king1997simpler}) for answering path maximum queries in a forest.

\begin{lemma}[\cite{komlos1985linear}] \label{lem:path-query}
Given a forest $F$ on $n$ vertices and $q$ query paths specified by their endpoints, the maximum edge weight on each query path can be determined deterministically using $O(n+q)$ comparisons.
\end{lemma}

\section{The Certificate Procedure}\label{sec:ceritificateAlgorithm}

\subsection{Overview and Formal Algorithm}

The entire $\Cert$ procedure is predicated around finding cycles of edges and deleting their heaviest edges. As mentioned in the technical overview, the algorithm seeks to iteratively \emph{project} vertices to smaller and smaller vertex sets to aid in cycle discovery. To this end, the algorithm maintains a subset $P$ of the vertices, called \emph{portals}. For each vertex $v$, it may maintain an incident edge $\sigma(v)$ connecting $v$ to a portal, called the \emph{spoke} of $v$. Let $\pi(v)$ denote the portal at the other endpoint of the spoke; if $v$ has no spoke, we write $\sigma(v)=\pi(v)=\,\perp$. 
For an edge $e=(u,v)$ whose endpoints both have spokes, the algorithm may \emph{project} $e$ to an edge of weight $w(e)$ with endpoints $\pi(u)$ and $\pi(v)$ in a child instance. Let $p(e)=\max\{w(\sigma(u)),w(\sigma(v))\}$ be the larger of the two spoke weights; we say the projection is \emph{valid} if $p(e)\le w(e)$.

The input to the certificate procedure is an undirected multigraph $G$. Each edge $e\in E(G)$ has a weight $w(e)$ and a \emph{projection record}, which stores, for each preceding recursion level, the pair of spokes used to project $e$ at that level. We say that $e$ is \emph{admissible} if every spoke in its projection record has weight at most $w(e)$, or equivalently, if all projections of $e$ at preceding recursion levels are valid. 
The admissibility test scans the projection record from the root toward the current call and reports the first recursion level at which it encounters an invalid projection, if any.

The certificate procedure maintains a forest $F\subseteq E(G)$ consisting of admissible edges and a set $R$ of rejected edges throughout its entire execution. The forest and rejected edges (as we will see later) get updated after each round is completed.  An edge is \emph{rejected} if it fails the admissibility test. Such edges are propagated upward until they reach the recursion level at which their first invalid projection was performed.

An edge $e=(u,v)$ is \emph{supported} by $F$ if $F_{\leq w(e)}$ connects $u$ and $v$. As we will show, $F$ together with the unsupported edges contains an MSF of $G$. We may therefore filter out (discard) every supported edge outside $F$.  Thus, the goal of the certificate procedure is to construct a forest $F$ using only $O(m)$ comparisons that supports most of the edges. 
The following lemma bounds the number of retained edges, the total size of the edge sets passed to exact (recursive) MSF calls, and the number of weight comparisons performed outside those calls.

\begin{lemma}\label{lem:certificate}
Let $G$ be an undirected multigraph with $n$ vertices, $m$ edges, and no isolated vertices. For every fixed constant $\varepsilon>0$, there exists a deterministic certificate procedure that outputs a forest $F\subseteq E(G)$. The procedure may invoke exact MSF calls on edge sets $E_1,\ldots,E_\ell$, each defined on a subset of $V(G)$. Apart from these calls, the procedure performs $O_\varepsilon(m)$ weight comparisons. Let $U(F)\subseteq E(G)$ denote the set of edges unsupported by $F$. The procedure guarantees that
\[
|F|+|U(F)|+\sum_{i=1}^{\ell}|E_i| \leq \varepsilon m+C_\varepsilon\sqrt{nm},
\]
where $C_\varepsilon$ is a constant depending only on $\varepsilon$.
\end{lemma}

The procedure is outlined in \cref{alg:certificate}. For a call at recursion level $d$, we set $t=\gamma\cdot d^3,r=8t^3$, where $\gamma$ is a fixed constant. The algorithm consists of $t^2$ rounds, partitioned into $t$ blocks of $t$ rounds each. After initialization, we schedule each edge of $G$ to take action in \emph{exactly one round}.

Let $A$ be the set of edges scheduled for the current round. Each edge $e=(u,v)\in A$ takes one of two actions. The $\textsf{find}$ action searches for a valid projection, while the $\textsf{project}$ action projects $e$ to $(\pi(u),\pi(v))$ using the current spokes, regardless of whether the projection is valid. Only edges taking the $\textsf{project}$ action enter a child instance (i.e., a recursive $\Cert$ call). Ideally, we would find a valid projection for each edge before projecting it. However, even checking validity of the current projections for $k$ edges can take $\Theta(k)$ weight comparisons.
At a level handling $\Theta(m)$ edges, these checks alone can therefore cost $\Theta(m)$ comparisons per level.

At the start of each round, the \emph{preferred action} of an edge $e=(u,v)$ is $\textsf{find}$ if either endpoint has no spoke or if both spokes lead to the same portal. When the spokes lead to distinct portals, the preferred action is $\textsf{find}$ if $F_{\leq p(e)}$ connects $u$ and $v$, and $\textsf{project}$ otherwise. This preceding preference rule will be incredibly important in the bounding the number of unsupported edges.
Unfortunately, exactly computing the preferred action of each edge will be prohibitively expensive, and thus the final algorithm \emph{estimates} the preferred action of each edge. Its estimate is always $\textsf{find}$ when a spoke is missing or both spokes lead to the same portal. We will later show that the estimate is still correct for \emph{most
edges} whose spokes lead to distinct portals.

Within each block (i.e., a sequence of $t$ rounds) we also maintain a \emph{batch} $B$ of edges, initialized to be the empty set. Edges whose estimated preferred action is $\textsf{find}$ are collected in this batch $B$ and processed at the end of the block. Each edge $e=(u,v)$ whose estimated preferred action is \textsf{project} is instead projected to $(\pi(u),\pi(v))$, forming a graph $G'$ on the portal set $P$. We recursively invoke the certificate procedure on $G'$, obtaining a forest $F'$ and a set of rejected edges $R'$. Because $F'$ and $R'$ are projections of edges in $G$, we first lift them back to their corresponding edges in $G$. We further let $R''\subseteq R'$ denote the edges whose first invalid projection occurs at the \emph{current recursion level}. We then incorporate $F'$ and $R''$ into $F$ via an exact MSF call. The remaining rejected edges $R'\setminus R''$ must have had their first invalid projections at earlier recursive levels, and thus are added to $R$ for propagation to those levels.

At the end of each block, we process the batch $B$. This procedure identifies a set $D$ of \emph{declared edges} to be incorporated into $F$ via an exact MSF call, introduces a set of new portals, and updates the spokes so that every nonrejected edge in $B$ is either supported by $F$ or has a valid projection. Any edges found to be inadmissible during this process are added to $R$.

\begin{algorithm}
\caption{\Cert($G$)} \label{alg:certificate}
\If{$E(G)=\varnothing$}{\Return{$(\varnothing, \varnothing)$}}
$F\gets \varnothing, R\gets \varnothing, P\gets \varnothing$.\\
Initialize $\sigma(v)\gets \,\perp$, $\pi(v)\gets\,\perp$ for every $v\in V(G)$.\\
Schedule each edge of $G$ to take action in one of the $t^2$ rounds. \label{line:schedule}\\
\For{$t$ blocks} {
    $B\gets \varnothing$.\\
    \For{$t$ rounds} {
        Let $A$ be the set of edges scheduled for this round.\\
        Determine (estimate) the preferred action for each edge in $A$.\\
        Add all edges whose preferred action is $\mathsf{find}$ to the batch $B$.\\
        Initialize an empty graph $G'$ on vertex set $P$.\\
        \ForEach{$e=(u,v)$ whose preferred action is $\mathsf{project}$}{
            Add the projected edge $(\pi(u),\pi(v))$ to $G'$
            with weight $w(e)$.\\
            Append the spokes of $u$ and $v$ to the projection
            record.
        }
        $F',R'\gets\Cert(G')$.\\
        Lift the edges of $F',R'$ back to their corresponding edges in $G$.\\
        Let $R''\subseteq R'$ be the set of edges whose first invalid projection occurs at this level.\\
        $F\gets \MSF(G[F\cup F'\cup R''])$, $R\gets R\cup (R'\setminus R'')$.
    }
    Process the batch $B$, obtaining a set $D$ of declared edges, a set $P'$ of new portals, and a set $R'$ of new rejected edges. Spokes are also updated.\\
    $F\gets \MSF(G[F\cup D])$, $P\gets P\cup P', R\gets R\cup R'$.
}
\Return $(F,R)$.
\end{algorithm}

\subsection{The Batch Processing} \label{sec:batch}

We first describe how to process a batch $B$ of edges, then give an efficient implementation that bounds the number of weight comparisons and admissibility tests. 

For each vertex $v$, group the incident edges of $B$ by their other endpoint (this is only necessary because there may be parallel edges from $v$ to the same vertex). We define the representative of a group to be its \emph{minimum weight} admissible edge, if one exists. At $v$, select the $k=r^2$ lightest representatives, or all representatives if there are fewer than $k$. Let $S_v$ be the set of neighbors of $v$ reached by the selected edges. The implementation tests admissibility of edges as needed. Every selected edge passes the test, and every edge that fails the test is rejected. 
When an edge fails an admissibility test, we store its first invalid projection level and \emph{reuse} the recorded failure on every later test request for that edge in the batch.
We \emph{declare} the selected edges, adding these to a set $D$. In this way, $D$ becomes the union of declared edges over all vertices. As explained in \cref{alg:certificate}, we then replace $F$ by an MSF of $F\cup D$ via an exact MSF call.

We next select a set $P'$ of portals that intersects every $S_v$ of size $k$. Indeed, to do this, we call each such set $S_v$ \emph{covered} once it contains a selected portal. We find a small portal set $P'$ as follows: while some vertex belongs to at least $r$ uncovered sets, select it as a portal. Each selection covers at least $r$ additional sets, so this step selects at most $n/r$ portals. At termination, each vertex belongs to fewer than $r$ uncovered sets. Thus the remaining sets have fewer than $nr$ incidences in total. Each has size $k=r^2$, so fewer than $n/r$ sets remain. Select one vertex from each set that is still uncovered. Altogether, this means that $|P'|\leq 2n/r$. We then update the existing portal set $P$ by adding $P'$ to $P$.

For each $v$ with $|S_v|=k$, let $x$ be the first portal selected that belongs to $S_v$. The selected edge from $v$ to $x$ is the new \emph{candidate spoke} of $v$. If $v$ has no spoke, we assign it this candidate spoke. Otherwise, $v$ chooses the lighter of the candidate spoke and its current spoke $\sigma(v)$.

We obtain the following guarantee:

\begin{lemma}\label{lem:batch}
Let $B$ be a batch of edges of a graph $G$ on $n$ vertices. The batch procedure declares a set $D\subseteq B$ of at most $kn=r^2n$ admissible edges and selects a set $P'$ of at most $2n/r$ portals. Next, let $R'\subseteq B$ denote the edges that fail an admissibility test during the batch. After replacing $F$ by $\MSF(G[F\cup D])$, every edge $e=(u,v)\in B\setminus R'$
satisfies at least one of the following conditions.
\begin{itemize}
    \item The edge $e$ is supported by $F$, that is, $F_{\leq w(e)}$
          connects $u$ and $v$.
    \item Both endpoints have spokes and
          $p(e)=\max\{w(\sigma(u)),w(\sigma(v))\}\leq w(e)$, so $e$
          admits a valid projection.
\end{itemize}
\end{lemma}

\begin{proof}
Each vertex selects at most $k$ edges, so $|D|\leq kn$. The portal bound $|P'| \le 2n/r$ follows from the covering argument just discussed above. 

We define $R'$ to be the set of edges that failed admissibility test during the selection procedure. Now consider any edge $e=(u,v)\in B\setminus R'$. Note that $e$ may or may not pass the admissibility test, but any untested edge can be made admissible without changing the execution or output. If $u\in S_v$ or $v\in S_u$, then one endpoint selected the representative of the group containing $e$. That representative joins $u$ and $v$ and has weight at most $w(e)$. Since it belongs to $D$, \cref{lem:threshold} implies that the updated forest connects $u$ and $v$ in $F_{\leq w(e)}$.

Otherwise, neither endpoint selected the group containing $e$. It follows that each endpoint selected $k$ representatives, all of weight at most $w(e)$. Thus $|S_u|=|S_v|=k$, and each endpoint receives a candidate spoke of weight at most $w(e)$. The spoke update applied at the end can only improve on this bound, so $p(e)\leq w(e)$.
\end{proof}

\paragraph{Efficient implementation.} For each vertex $v$, consider its incident edges in $B$, grouped by their other endpoint. Our goal is to select the $k=r^2$ lightest edges among the minimum weigh admissible edges of these groups. We now describe how to implement this efficiently without performing an admissibility test on every edge. We bound the number of successful admissibility tests and failed tests will be charged separately.

\begin{lemma}\label{lem:batch-comparison}
Suppose $m$ edges of unknown admissibility are partitioned into groups. For each group containing an admissible edge, let its representative be its minimum weight admissible edge. Given an integer $k\geq 0$, we can select the $k$ lightest representatives, or all representatives if there are fewer than $k$, using $O(k\log(m+1))$ successful admissibility tests and $O(m)$ additional weight comparisons.
\end{lemma}

\begin{proof}
We use a recursive procedure. If $k=0$, return the empty set. If $m\leq k$, test every edge and return the minimum weight admissible edge from each group that has one.

Otherwise, use deterministic median selection~\cite{BFP+72} to partition the edges into a lower half $L$ of size $\lfloor m/2\rfloor$ and an upper half $H$, using the fixed edge order to break weight ties. Scan $L$, testing edges until either $k$ distinct groups have passed a test or the scan is exhausted. Once an edge passes, skip the remaining edges of its group for the rest of this scan.

If $k$ groups pass, at least $k$ groups have representatives in $L$, so all $k$ desired representatives lie in $L$. We then recurse on $L$ with quota $k$.

Otherwise, let $k' <k$ be the number of groups that passed. These are exactly the groups with an admissible edge in $L$, and each contributes a representative to the answer. For each such group, recurse with quota one on its edges in $L$ to find that representative. Remove all edges of these $k'$ groups from $H$, and recurse on the remaining edges of $H$ with quota $k-k'$. Every remaining representative must come from this last call.

The correctness follows immediately. We now analyze the number of comparisons.  Deterministic median selection requires $O(m)$ comparisons (see \cite{BFP+72}). First consider a call with quota one. Such a call never starts another quota-1 group search. At each level, it performs $O(m)$ comparisons and at most one successful admissibility test, and recurses on half of the edges. Therefore, the call uses $O(m)$ comparisons and $O(\log (m ))$ successful tests in total.

Now consider an arbitrary call with quota $k$. At each level, the procedure performs $O(m)$ comparisons and at most $k$ successful admissibility tests. Moreover, there is at most one recursive call with quota more than one, and the call contains at most half of the edges. All other recursive calls are quota-1 group searches which operate on disjoint subsets of the discarded half. Therefore, the main recursion uses $O(m)$ comparisons and $O(k \log (m))$ successful tests. The quota-$1$ group searches contribute another $O(m)$ comparisons since their inputs are disjoint, and $O(k\log (m))$ successful tests since there are at most $k$ group searches in total. This completes the proof.
\end{proof}

We apply \cref{lem:batch-comparison} to the edges incident to each vertex $v$ with quota $k$. The total number of comparisons is therefore bounded by $\sum_{v} O(\deg_B(v)) = O(|B|)$. In addition, the total number of successful admissibility tests is at most
\[
\sum_v O(k\log (\deg_B(v)+1)) \leq \sum_v O\left(k \sqrt{\deg_B(v)}\right) = O\left(r^2 \sqrt{|B|\cdot n}\right),
\]
where we used $\log(x+1)\leq \sqrt{x}$, Cauchy--Schwarz, and the fact that $k=r^2$.

Throughout batch processing, comparisons other than admissibility tests arise from two sources: $O(|B|)$ from the bound above and $O(n)$ comparisons for updating the spokes, yielding a total of $O(n+|B|)$ comparisons.

\subsection{Bounding Unsupported Edges}
Recall that an edge $e=(u,v)$ is supported by $F$ if $F_{\leq w(e)}$ connects $u$ and $v$. Our goal is to ensure that most edges are supported by the final forest $F$. In this section, we characterize the unsupported edges and bound their number.

Fix an edge $e=(u,v)\in E(G)$. For the analysis, we consider its preferred action at the start of each round, even though each edge is assigned to take action in only one round. Recall that its preferred action is $\textsf{find}$ if (1) either endpoint has no spoke, (2) $\pi(u)=\pi(v)$, or (3) $F_{\le p(e)}$ connects $u,v$; otherwise, it is $\textsf{project}$.

Recall that the $t^2$ rounds are partitioned into $t$ blocks of $t$ rounds each. Within each block, the spokes remain unchanged; they are updated only at the end of the block when a batch is processed. In addition, the connectivity in $F$ only improves as the rounds progress -- that is to say, connectivity in $F_{\leq a}$ is preserved at every possible choice of the threshold $a$. This means that whenever a spoke $s$ is declared, going forward, it will always be the case that $F_{\leq w(s)}$ connects its endpoints. Thus, within each block, the preferred action of $e$ is \textsf{project} for an initial sequence of rounds, followed by \textsf{find} for the remaining rounds, where either sequence may be empty.

We now define the notion of a \emph{bad} round for $e$; all other rounds are called \emph{good}.

\begin{definition}
A round is \emph{bad} for $e$ if either of the following holds:
\begin{itemize}
    \item It is the last round in its block with a \textsf{project} preference.
    \item The preferred action is \textsf{find}, and the block is the last block containing any \textsf{find} preference.
\end{itemize}
\end{definition}

By definition, there are at most $2t$ bad rounds. Indeed, there are at most $t$ blocks, each containing at most one bad \textsf{project} round. In addition, bad \textsf{find} rounds can occur in at most one block, which contains at most $t$ rounds.

Recall that the algorithm does not compute the true preferred action for every edge, but instead uses estimates. We say that an \emph{error} occurs for an edge $e$ if either the schedule assigns $e$ to take action in a bad round, or the estimated preferred action in its assigned round differs from the true preferred action. We establish the following key property: if no error occurs for $e$ throughout the recursion of \cref{alg:certificate}, then $e$ is supported by the final forest $F$.

To prove this, we first establish the following useful properties.

\begin{claim} \label{clm:admissible}
For a call $\Cert(G)$, every edge in the returned forest $F$ is admissible.
\end{claim}
\begin{proof}
    We consider all edges that may be inserted into $F$. Every declared edge in $D$ passes the admissibility test during batch processing, as described in \cref{sec:batch}. All the returned child forest edges are admissible by induction. A rejected edge is inserted at the first recursion level where it performs an invalid projection, and hence it is admissible before that projection. Therefore, every edge inserted into $F$ is admissible.
\end{proof}

\begin{claim} \label{clm:child-path}
For an edge $(u,v)$ with projection $(\pi(u),\pi(v))$, if the child recursion supports the projected edge, then on return $F_{\leq \max\{w(e),p(e)\}}$ connects $u$ and $v$.
\end{claim}
\begin{proof}
Since the projected edge is supported, there exists a path in the child forest connecting $\pi(u)$ and $\pi(v)$ using only edges of weight at most $w(e)$. Consider every edge $f'=(\pi(x),\pi(y))$ on this path, note that $f'$ is first lifted back to the corresponding edge $f=(x,y)$ in $G$ before being incorporated into $F$. By \cref{clm:admissible}, $f'$ is admissible, so both spokes $\sigma(x)$ and $\sigma(y)$ used in this lifting have weight at most $w(f)\leq w(e)$. Thus, replacing each edge $f'$ on the child forest path with the walk $(\pi(x), x, y, \pi(y))$ using $\sigma(x)$ the lifted edge of $f'$ and $\sigma(y)$. Concatenating these walks and adding $\sigma(u)$ and $\sigma(v)$ gives a walk from $u$ to $v$ with all weights at most $W=\max\{w(e),p(e)\}$. By the preceding arguments, every edge on this path has weight at most $w(e)$, except for $\sigma(u),\sigma(v)$. Moreover, these edges are either spokes or child forest edges, which are all incorporated into $F$ via exact MSF calls (note that spokes can only arise from declared edges $D$ during batch processing). 
By \cref{lem:threshold}, $u$ and $v$ are connected in $F_{\leq \max\{w(e),p(e)\}}$.
\end{proof}

Now we are ready to prove the key lemma, which shows that every unsupported or rejected edge can be charged to an error.

\begin{lemma} \label{lem:error}
Consider a root instance $\Cert(G)$ and an edge $e\in E(G)$. Suppose no error occurs for $e$ or any of its projections throughout the recursion, then $e$ is supported by the final forest $F$ returned by the algorithm. If a rejected edge has its first invalid projection at some recursion level, then an error occurs for that edge at that level.
\end{lemma}
\begin{proof}
Fix an edge $e\in E(G)$, and suppose that throughout the recursion, no error occurs for $e$ or any of its projections, none of them is rejected, and all their estimated preferred actions are correct. Note that an edge is projected and passed to exactly one child recursion instance if its preferred action at its assigned round is \textsf{project}; if its preferred action is $\textsf{find}$, it does not appear in any subsequent recursion instance. We prove by backward induction on the recursion level that the projection $e'=(u',v')$ of $e$ at every level is supported.

Consider the last recursion level in which the projection of $e$ exists. Its preferred action must be $\textsf{find}$. Thus, $e'$ is included in the batch and processed at the end of the block. By \cref{lem:batch}, either $e'$ is supported, or its spokes satisfy $p(e')\leq w(e')$. In the former case, we are done. In the latter case, since the assigned round is good, there must be a later block in which $e'$ prefers \textsf{find}. In that round, either the spokes coincide ($\pi(u')=\pi(v')$), or $F_{\leq p(e')}$ connects $u',v'$. Since $p(e')$ can only decrease, we still have $p(e')\leq w(e')$. In the former case, the two spokes meet at a common portal, so $F_{\leq p(e')}$ connects $u'$ and $v'$. Therefore, in either case, $F_{\leq p(e')}$ and hence $F_{\leq w(e')}$ connects $u',v'$, and $e'$ is supported.

Now consider a recursion level where the preferred action of $e'$ is \textsf{project}. In its assigned round, $e'$ is projected to $(\pi(u'),\pi(v'))$ and passed to a child recursion instance. By \cref{clm:child-path}, on return, $F_{\leq \max\{w(e'),p(e')\}}$ connects $u'$ and $v'$. If $p(e')\leq w(e')$, $e'$ is supported as desired. Otherwise, $F_{\leq p(e')}$ connects $u'$ and $v'$. However, since the assigned round is good, the preferred action in the next round must also be \textsf{project}, contradicting the fact $F_{\leq p(e')}$ connects $u',v'$. Therefore, $e'$ must be supported. The first claim follows by backward induction up to the root recursion instance.

For a rejected edge $e$, suppose an admissibility test fails and identifies the first level where an invalid projection occurs. When $e$ is projected in some round at this level, we have $p(e)>w(e)$. At the end of that round, the edge returns from the recursive call and is incorporated into the forest $F$ via an exact MSF call. By \cref{lem:threshold}, $u$ and $v$ are connected in $F_{\leq p(e)}$ at that point. Since the spokes remain unchanged in the block containing this round, $p(e)$ remains fixed, and $u$ and $v$ remain connected in $F_{\leq p(e)}$. Thus, all subsequent rounds in the block prefer \textsf{find}. If the estimated preferred action is correct, the assigned round must be bad. Otherwise, the estimate is incorrect. In either case, an error occurs. This
also proves the first claim for a rejected edge.
\end{proof}

\subsection{Building the Schedule}
The preceding sections have mostly dealt with the complexity of ensuring the algorithm behaves well, provided the edges are scheduled to their rounds appropriately.
In this section, we now establish a deterministic schedule that assigns most edges to good rounds. There is a subtle circularity: the preferred actions of edges depend on the schedule, which in turn determines which rounds are good or bad. To reason about this dependence, we introduce an object called the \emph{history}.

\begin{definition}[History]
Consider a call $\Cert(G)$ on a graph $G$ with $n$ vertices and $m$ edges. We index the $t^2$ rounds by $i\in [t^2]$. We define the history to be
\[
H=(\sigma_i,F_i,a_i)_{i\in [t^2]},
\]
where:
\begin{itemize}
    \item $\sigma_i$ is a vector of length $n$, where $\sigma_i(v)\in E(G)\cup{\perp}$ specifies the spoke of vertex $v$ at the start of round $i$.
    \item $F_i\subseteq E(G)$ is the forest $F$ maintained by the algorithm at the start of round $i$.
    \item $a_i$ is a vector of length $m$, where $a_i(e)\in \{\textsf{find},\textsf{project}\}$ specifies the \emph{true preferred action} of edge $e$ at the start of round $i$.
\end{itemize}
\end{definition}

Note that the history completely determines which rounds are good or bad for each edge. Ultimately, we will show that the number of possible histories is not too large. This will allow us to show that there exists a deterministic schedule that satisfies the desired property for every possible history of the algorithm.

\paragraph{Bounding the number of histories.}

We use the following bound on the number of possible vectors of maxima, appearing in \cite{graham1980information} attributed to Fredman. We include a proof for completeness.

\begin{lemma}\label{thm:maximizingVectors}
Let $S_1,\dots,S_m$ be arbitrary nonempty subsets of a ground set of size $n$.  For each total order $\pi$ over the ground set, let $x_{\pi}$ be the vector that records the names of the maxima for each set $S_i$. Then, as $\pi$ ranges over all possible total orders, the number of distinct vectors $x_{\pi}$ is at most $n+m-1\choose m$.
\end{lemma}
\begin{proof}
For each total order $\pi$, let $y_\pi$ record, for each element, the number of sets in which it is the maximum. Since $\sum_{j}y_\pi(j)=m$, there are at most $\binom{n+m-1}{m}$ possible vectors $y_\pi$. We claim that $y_\pi$ uniquely determines $x_\pi$.

Suppose otherwise that there exist two total orders $\pi_1,\pi_2$ such that $y_{\pi_1}=y_{\pi_2}$ but $x_{\pi_1}\neq x_{\pi_2}$. Construct a directed multigraph as follows: for each set $S_i$ whose maxima under $\pi_1$ and $\pi_2$ differ, add a directed edge from its maximum under $\pi_2$ to its maximum under $\pi_1$. Thus, every edge points to a strictly larger element in $\pi_1$. However, $y_{\pi_1}=y_{\pi_2}$ implies that every element has equal in-degree and out-degree. Since the graph is nonempty, it must contain a directed cycle, a contradiction
\end{proof}

With this in hand, we now bound the number of possible histories.

\begin{lemma}\label{lem:historyBound}
Consider a call $\Cert(G)$ on a graph $G$ with $n$ vertices and $m$ edges. Let $\mathcal{H}$ denote the set of all possible histories of this call, obtained by enumerating every sequence of possible spokes and forests that appear at each round, and then enumerating every possible list of preferences induced by these spokes and forests. Then
\[
    \log |\mathcal{H}| = O(t^2\sqrt{mn}).
\]
\end{lemma}

\begin{proof}
    First we bound the number of possible sequences $(\sigma_i)_{i\in[t^2]}$. In each round, the spoke $\sigma_i(v)$ has at most $\deg_G(v)+1$ possible values. Thus, the total number of such sequences is at most
    \[
    \left(\prod_{v \in V} (\mathrm{deg}(v)+1)\right)^{t^2}.
    \]
    Taking logarithms, we obtain
    \begin{align*}
        \log \left(\prod_{v \in V(G)} (\mathrm{deg}(v)+1)\right)^{t^2} & = t^2\cdot \sum_{v\in V(G)} \log (\deg_G(v)+1) &&\\
        & \leq t^2 \cdot \sum_{v\in V(G)} \sqrt{\deg_G(v)} &&(\log(x+1)\leq \sqrt{x})\\
        & \leq t^2 \sqrt{2mn}. && (\text{Cauchy--Schwarz})
    \end{align*}

    Next, we bound the number of possible sequences $(F_{i})_{i\in [t^2]}$. Since each forest uses at most $n-1$ edges out of the $m$ edges in the graph, the total number of possible sequences is at most
    \[
    \binom{m + n-1}{n-1} ^{t^2}.
    \]
    Taking logarithms, we obtain
    \[
    \log \binom{m + n-1}{n-1} ^{t^2} \leq t^2\log \binom{m+n}{m} \leq 2t^2\sqrt{mn},
    \]
    where we used \cref{fact:binom}.

    Finally, we bound the number of possible true preference vectors $(a_i)_{i\in [t^2]}$. Fix the spokes $\sigma_i$ and the forest $F_i$ at the start of round $i$. For each edge $e=(u,v)\in E(G)$, recall that its preferred action is $\textsf{find}$ if (1) either endpoint has no spoke, (2) $\pi(u)=\pi(v)$, or (3) $F_{\le p(e)}$ connects $u,v$; otherwise, it is $\textsf{project}$. The first two cases are completely determined by the spokes. For any remaining edges that are not connected in $F_i$, their action is forced to be $\textsf{project}$. So it remains only to analyze how edges in the same component that are also in case (3) can be assigned to $\textsf{project}$ and $\textsf{find}$.

    Consider a ground set $T$ of size at most $2n$ consisting of all the spokes and edges of $F_i$. The total order is determined by edge weights, with spokes ranked higher than forest edges in case of ties. For each edge $e=(u,v)\in E(G)$, let $S_e\subseteq T$ consist of the spokes $\sigma_i(u),\sigma_i(v)$ and all edges on the $u$ to $v$ path in $F_i$, if such a path exists. The preferred action of $e$ is \textsf{find} precisely when the maximum element of the set $S_e$ is a spoke (either $\sigma_i(u)$ or $\sigma_i(v))$. Indeed, this occurs precisely when the heaviest spoke is at least as heavy as every edge on the $u$ to $v$ path in $F_i$, which is equivalent to $u,v$ being connected in $(F_i)_{\leq p(e)}$. Conversely, if the maximum weight edge in $S_e$ is a path edge (weight strictly larger than both spokes), then $u$ and $v$ are not connected in $(F_i)_{\leq p(e)}$.  Note that here we use the fact that there is a \emph{unique} path between $u$ and $v$ in $F_i$ (and that $u,v$ are connected - as disconnected $u,v$ are already determined by the previous discussion). 

    This now exactly mirrors the setting of \cref{thm:maximizingVectors} on a ground set of size at most $2n$ with $m$ sets. Consequently, for any fixed choices of the spokes and forests, the number of possible true preference vectors is at most
    \[
    \binom{m+2n-1}{m}^{t^2}.
    \]
    Taking logarithms, we obtain
    \[
    \log \binom{m+2n-1}{m}^{t^2} \leq t^2\log \binom{m+2n}{m} \leq 2t^2\sqrt{2mn},
    \]
    where we used \cref{fact:binom}.
    
    Combining the three bounds, we obtain
    \[
    \ln \left | \mathcal{H}\right |  \leq t^2\sqrt{2mn}+2t^2\sqrt{mn}+2t^2\sqrt{2mn} = O(t^2\sqrt{mn})
    \]
    as desired.
\end{proof}

\paragraph{Finding a good schedule.} We next show that, at the start of the algorithm, we can fix a deterministic schedule under which most edges perform their actions in good rounds for every possible history.

\begin{lemma} \label{lem:schedule}
Consider a call $\Cert(G)$ on a graph $G$ with $n$ vertices and $m$ edges. For each history $H$ and schedule $\rho: E(G) \rightarrow [t^2]$, we let $X_H(\rho)$ denote the number of edges whose action is scheduled in one of their bad rounds. Then there exists a schedule $\rho^*$ such that for every $H \in \mathcal{H}$,
    \[
    X_H(\rho^*) \leq \frac{\log |\mathcal{H}| + 2m/t}{\log 2}.
    \]
\end{lemma}

\begin{proof}
    Fix an arbitrary history $H\in\mathcal{H}$. Recall that this determines the good and bad rounds for each edge. Now consider a random schedule $\rho$ that independently assigns each edge to a uniformly random round. For each edge $e\in E(G)$, let $Y_e$ be a random variable with $Y_e=1$ if $e$ is assigned to a bad round, and $Y_e=0$ otherwise. Since each edge has at most $2t$ bad rounds out of $t^2$ rounds, we have
    \[
    \E_{\rho}\left[2^{Y_e}\right] = 1+\Pr[Y_e=1] \leq 1+\frac{2t}{t^2} = 1 + \frac{2}{t}.
    \]
    Since $X_H(\rho)=\sum_{e\in E(G)} Y_e$, and the assignments are independent, it follows that
    \[
    \E_{\rho}\left[2^{X_H(\rho)}\right] = \prod_{e\in E(G)} \E_{\rho}\left[2^{Y_e}\right] \leq \left(1+\frac{2}{t}\right)^m \leq e^{2m/t}.
    \]
    Summing over all histories gives
    \[
    \E_{\rho}\left[\sum_{H\in \mathcal{H}}2^{X_H(\rho)}\right] \leq |\mathcal{H}| \cdot  e^{2m/t}.
    \]
    By averaging, this implies that there exists a schedule $\rho^*$ such that
    \[
        \sum_{H\in\mathcal{H}}2^{X_H(\rho^*)}
        \leq |\mathcal{H}| \cdot e^{2m/t}.
    \]
    In particular, for every $H\in\mathcal{H}$,
    \[
    2^{X_H(\rho^*)} \leq |\mathcal{H}| \cdot  e^{2m/t},
    \]
    and hence
    \[
    X_H(\rho^*) \leq \frac{\log |\mathcal{H}|+2m/t}{\log2}
    \]

\end{proof}

The key observation is that this allows us to choose a schedule $\rho^*$ \emph{independent} of the actual execution of the certificate call. In particular, we can enumerate all possible schedules and check all possible histories to find a good schedule. Although this may incur enormous computational overhead, it requires no comparisons, and we are only concerned with the number of comparisons made by the algorithm.

\subsection{Estimating Preferred Actions} \label{sec:estimate}
In the preceding section, we described how to choose a schedule that determines the round in which each edge takes action. In this section, we show how to efficiently estimate the preferred actions, ensuring that the estimates are correct for most edges.

Consider a call $\Cert(G)$ on a graph $G$ with $n$ vertices and $m$ edges. Fix a round $i$, and let $A_i$ denote the set of edges assigned to this round, with $m_i=|A_i|$. Consider the vector $a_i$ of length $m_i$, which specifies the preferred action for each edge in $A_i$. By the argument in \cref{lem:historyBound} (via \cref{thm:maximizingVectors}), conditioned on the current forest $F_i$ and spokes $\sigma_i$, the number of possible preference vectors is at most $e^{O(\sqrt{m_i n})}$. Let $\mathcal A$ denote the family of all such vectors.

We first show that, by querying the true preferred actions of a small subset of edges, we can identify a vector in $\mathcal A$ that differs from $a_i$ on at most $m_i/t$ coordinates.

\begin{claim}
There exists a subset $S^*\subseteq A_i$ of size $O(t\sqrt{m_i n})$ such that any two vectors $a,b\in\mathcal A$ that agree on all coordinates corresponding to $S^*$ differ in at most $m_i/t$ coordinates.
\end{claim}
\begin{proof}
We sample $k=2 t(\log |\mathcal{A}|+1)$ edges from $A_i$ independently and uniformly at random. Let $S$ be the set of sampled edges.

Fix any two vectors $a,b\in\mathcal{A}$ that differ in more than $m_i/t$ coordinates. The probability that they agree on all coordinates corresponding to $S$ is at most $(1-1/t)^k \leq e^{-k/t}$. By a union bound over all pairs $a,b\in\mathcal A$, the probability that there exist two vectors differing in more than $m_i/t$ coordinates but agreeing on all coordinates in $S$ is at most $|\mathcal{A}|^2 e^{-k/t} <1$. Therefore, there exists a choice $S^*$ satisfying the desired property. Since $|\mathcal{A}|= e^{O(\sqrt{m_in})}$, we have $|S^*|=O(t\sqrt{m_in})$.
\end{proof}

Again, the algorithm can explicitly find such a set $S^*$ by enumerating and checking, without making any comparisons. It then queries the true preferred actions for edges in $S^*$ and selects any vector in $\mathcal{A}$ consistent with the queried answers as the estimate. These queries correspond precisely to path maximum queries on the forest $F$, which can be answered using the algorithm from \cref{lem:path-query}. In round $i$, there are $O(t\sqrt{m_i n})$ queries, and it requires $O(n+t\sqrt{m_i n})$ comparisons. Summing over all $t^2$ rounds, the total number of comparisons is at most
\[
\sum_{i=1}^{t^2}O(n+t\sqrt{m_i n}) = O(t^2n+t^2\sqrt{mn})
\]
by Cauchy--Schwarz. In addition, the total number of edges with incorrect estimates is at most
\[
\sum_{i=1}^{t^2} \frac{m_i}{t} = \frac{m}{t}.
\]

\subsection{Putting the Pieces Together}
We now prove \cref{lem:certificate}. Recall that the parameters are set to $t=\gamma \cdot d^3, r=8t^3$, where $\gamma$ is a fixed constant. We also observe that $n\leq 2m$ since the initial graph has no isolated vertices.

For a call on $n$ vertices, the batch processing procedure is executed $t$ times, each introducing at most $2n/r$ new portals. Thus, the total number of portals is bounded by $2nt/r$. Since the algorithm makes $t^2$ recursive calls, the total number of vertices across all child instances is bounded by
\[
t^2 \cdot \frac{2nt}{r} = \frac{2nt^3}{r} = \frac{n}{4}
\]
by the parameter choices. Therefore, the total number of vertices decreases by a factor of 4 in each recursion level.

For each recursion level $d$, let $n_d$ be the total number of vertices and $m_d$ the total number of edges. By the above argument, we have $n_d\leq 4^{-(d-1)}\cdot n$ and $m_d\leq m$ since each edge is passed to at most one child instance. For each call $c$, let $d_c$ denote its recursion level, $n_c$ and $m_c$ its numbers of vertices and edges, and $t_c=\gamma \cdot d_c^{3},r_c=8t_c^3$ its parameter values. We obtain the following useful property.
\begin{claim}
\[
\sum_{\text{calls } c} \text{poly}(d_c,t_c, r_c) (n_c+\sqrt{m_cn_c}) = O_{\gamma}(\sqrt{mn}).
\]
\end{claim}
\begin{proof}
Fix a recursion level $d$, and consider all calls $c$ at level $d$, we have
\[
\sum_{c} \text{poly}(d_c,t_c,r_c) (n_c+\sqrt{m_cn_c}) \leq O_{\gamma}\left(\text{poly}(d) (n_d+\sqrt{m_dn_d})\right).
\]
by Cauchy--Schwarz and the fact that $t_c,r_c=\text{poly}(d_c,\gamma)$. Since $n_d\leq 4^{-(d-1)}\cdot n$ and $n\leq 2m$, this is at most
\[
O_{\gamma}\left(\frac{\text{poly}(d)}{2^{d-1}} \cdot \sqrt{mn}\right).
\]
Summing over all recursion levels, we obtain
\[
\sum_{\text{calls } c} \text{poly}(d_c) (n_c+\sqrt{m_cn_c}) \leq \sum_{d\geq 1} O_{\gamma}\left( \frac{\text{poly}(d)}{2^{d-1}} \cdot \sqrt{mn}\right) = O_\gamma(\sqrt{mn}).
\]
\end{proof}

\paragraph{Bounding the number of errors.}
For a call $c$, an error on an edge $e$ can arise from two sources: either the schedule assigns $e$ to take action in a bad round, or the estimated preferred action on its assigned round is incorrect. By \cref{lem:schedule,lem:historyBound}, the number of errors of the first type is at most
\[
\frac{\log |\mathcal{H}|+2m_c/t_c}{\log 2} \leq O(t_c^2\sqrt{m_cn_c}) + \frac{2m_c}{t_c\log2}.
\]
By the analysis in \cref{sec:estimate}, the number of errors of the second type is at most $m_c/t_c$. Therefore, the total number of errors in call $c$ is at most
\[
O(t_c^2\sqrt{m_cn_c}) + \left(1+\frac{2}{\log 2}\right) \frac{m_c}{t_c} \leq O(t_c^2\sqrt{m_cn_c}) +\frac{4m_c}{t_c}.
\]
Summing over all calls, we obtain
\[
\sum_{\text{calls } c} O(t_c^2\sqrt{m_cn_c}) +\frac{4m_c}{t_c} \leq O_\gamma(\sqrt{mn}) + \sum_{d\geq 1} \frac{4m_d}{\gamma\cdot d^3} \leq  O_\gamma(\sqrt{mn})+\frac{8m}{\gamma}.
\]

By \cref{lem:error}, the same bound applies to both the number of unsupported edges.

\paragraph{Bounding the number of comparisons.} We now bound the number of comparisons performed in a call $c$, excluding the exact MSF calls.

First, consider the comparisons used in batch processing. Let $B_1,\dots,B_{t_c}$ denote the batches. By the analysis in \cref{sec:batch}, processing a batch $B_i$ requires $O(n_c+|B_i|)$ comparisons. The $O(n_c)$ term contributes $O(t_cn_c)$ over the entire call. For the $O(|B_i|)$ term, since each edge enters at most one batch throughout the entire recursion, these comparisons contribute $O(m)$ in total across all calls. In addition, by the analysis in \cref{sec:batch}, processing $B_i$ requires $O(r_c^2\sqrt{|B_i|\cdot n_c})$ successful admissibility tests. Each successful test requires $O(d_c)$ comparisons. Therefore, the total number of comparisons due to successful tests is at most
\[
O(d_c)\cdot\sum_{i=1}^{t_c} O\left(r_c^2 \sqrt{|B_i|\cdot n_c} \right) = O\left(d_cr_c^2\sqrt{t_cm_cn_c}\right)
\]
by Cauchy--Schwarz.

Next, by the analysis in \cref{sec:estimate}, estimating the preferred actions requires $O(t_c^2 n_c+t_c^2\sqrt{m_cn_c})$ comparisons. Summing these bounds over all calls, we obtain
\[
\sum_{\text{calls }c} O\left(t_cn_c+d_cr_c^2\sqrt{t_cm_cn_c}+t_c^2 n_c+t_c^2\sqrt{m_cn_c}\right) = O_{\gamma}(\sqrt{mn}).
\]

Finally, consider the comparisons incurred by failed admissibility tests. Each failed test causes an edge to be rejected. If the first invalid projection of a rejected edge occurs at recursion level $d$, then the test costs at most $O(d)$ comparisons. By \cref{lem:error} and the preceding bound on the number of errors, the total number of comparisons due to failed tests is at most
\[
\sum_{\text{calls }c} O(d_c) \cdot \left(O(t_c^2\sqrt{m_cn_c})+\frac{4m_c}{t_c}\right) = O_{\gamma}(\sqrt{mn}) + O\left(\sum_{d\geq 1}\frac{4dm_d}{\gamma \cdot d^3} \right) = O_{\gamma}(\sqrt{mn}) + O(m).
\]

Combining the above bounds, the total number of comparisons is at most $O_{\gamma}(\sqrt{mn}) + O(m) = O(m)$ since $n\leq 2m$.

\paragraph{Bounding the total size of exact MSF calls.}
For a call $c$, the algorithm makes an exact MSF call at each of the $t_c^2$ rounds. Each invocation takes as input the old forest and the child forest, each containing at most $n_c-1$ edges, together with the rejected edges whose first invalid projection occurs at the current level. In addition, at the end of each block, the algorithm makes an exact MSF call on the old forest, containing at most $n_c-1$ edges, and the set of declared edges $D$ of size at most $r_c^2n_c$. Thus, across all calls where $m_c > 0$, the total input size of all exact MSF calls is at most
\[
[\text{\# rejected edges}]+ \sum_{\text{calls }c: m_c > 0} \left(2t_c^2(n_c-1) + t_c(n_c-1+r_c^2n_c)\right)
\]
By \cref{lem:error} and the preceding bound on the number of errors, the first term is at most $O_\gamma(\sqrt{mn})+8m/\gamma$. The second term is at most $O_\gamma(\sqrt{mn})$. Therefore, the total input size of all exact MSF calls is bounded by $O_\gamma(\sqrt{mn})+8m/\gamma$.

Overall, we have
\[
|F|+|U(F)|+\sum_{i=1}^{\ell} |E_{i}| \leq O_\gamma(\sqrt{mn}) + \frac{16m}{\gamma}.
\]
Taking $\gamma\geq 16/\varepsilon$ proves \cref{lem:certificate}.

\section{The Final Algorithm}

With the certificate algorithm established, we are ready to present and prove our main result.

\begin{theorem}\label{thm:main2}
    Let $\mathcal{T}^*(m,n)$ be the deterministic decision-tree complexity of the $\mathrm{MSF}$ problem on graphs with $m$ edges and $n$ vertices. Then $\mathcal{T}^*(m,n) = O(m)$.
\end{theorem}

Combined with the result of \cite{pettie2002optimal}, this immediately yields the following corollary.

\begin{corollary}
There exists a uniform deterministic $O(m)$ time algorithm for computing the MST of any connected graph with $m$ edges and $n$ vertices, in both the RAM model and pointer machine model. In general, there is a deterministic $O(m+n)$ time algorithm for computing the MSF of any graph (the extra $O(n)$ time is necessary for even reading over all of the vertices).
\end{corollary}

The corollary above proves \cref{thm:main}. We now prove \cref{thm:main2}.

\begin{algorithm}
    \caption{$\MSF(G)$} \label{alg:MSF}
    Perform $k$ Bor\r{u}vka steps on $G$ for a large enough constant $k$.\\
    Let $S$ be the set of contracted edges, and let $G'$ be the resulting contracted graph.\\
    Remove all isolated vertices from $G'$.\\
    \If{$G'=\varnothing$} {
        \Return $S$.
    }
    Run $\Cert(G')$ to obtain a forest $F\subseteq E(G')$.\\
    Compute the set $U(F)$ of edges unsupported by $F$.\\
    \Return $S\cup\MSF(G'[F\cup U(F)])$.
\end{algorithm}

\begin{proof}[Proof of \cref{thm:main2}]
Consider a graph $G$ with $n$ vertices and $m$ edges. We may assume without loss of generality that $G$ has no isolated vertices, and hence $n\leq 2m$. We also assume that all edge weights are distinct with ties broken by edge indices.

The algorithm is described in \cref{alg:MSF}. It first performs a constant number of Bor\r{u}vka steps to reduce the number of vertices, then invokes the certificate procedure from \cref{lem:certificate} to obtain a forest $F$, and finally recurses on $F\cup U(F)$ where $U(F)$ is the set of edges unsupported by $F$.

We set $\varepsilon=1/8$, and choose a sufficiently large constant integer $k$ such that $C_{1/8} \cdot 2^{(1-k)/2} \leq 3/8$, where $C_{1/8}$ is the constant from \cref{lem:certificate}.

The algorithm begins by performing $k$ Bor\r{u}vka steps. This requires $O(km)=O(m)$ comparisons. Let $S$ denote the set of edges contracted during these steps. The number of remaining non-isolated vertices is at most $n/2^k$. We then remove all isolated vertices from the contracted graph and let $G'$ denote the resulting graph, with $n'\leq n/2^k$ vertices. If $G'$ is empty, the algorithm returns $S$ and terminates. Otherwise, we apply the certificate procedure from \cref{lem:certificate} to $G'$.

The call $\Cert(G')$ returns a forest $F$ and makes recursive exact MSF calls on edge sets $E_1,\ldots,E_\ell$. Let $U(F)\subseteq E(G')$ denote the set of edges unsupported by $F$. By \cref{lem:certificate}, we have
\[
|F|+|U(F)|+\sum_{i=1}^{\ell}|E_i| \leq \frac{1}{8} m+C_{1/8}\sqrt{mn'}.
\]
Moreover, given the forest $F$ and $E(G')$, we can efficiently identify $U(F)$ using the algorithm from \cref{lem:path-query}. This requires only $O(m)$ additional comparisons.

By the correctness of the Bor\r{u}vka steps and the cycle property (\cref{fact:cycleProperty}), we know that
\[
\MSF(G) = S\cup \MSF(G'[F\cup U(F)]).
\]
Thus, the algorithm recursively computes $\MSF(G'[F\cup U(F)])$ and combines the result with $S$ to obtain $\MSF(G)$.

The total number of edges across all recursive MSF calls is bounded by
\begin{align*}
|F\cup U(F)|+\sum_{i=1}^{\ell} |E_i| & \leq \frac{m}{8} + C_{1/8} \sqrt{mn'} &&\\
& \leq \frac{m}{8}+\frac{C_{1/8}}{2^{k/2}} \sqrt{mn}  &&(n'\leq n/2^k)\\
& \leq \frac{m}{8} + \frac{C_{1/8}}{2^{k/2}} \sqrt{2m^2} &&(n\leq 2m)\\
& \leq \frac{m}{8}+\frac{3m}{8} && (C_{1/8}\cdot 2^{(1-k)/2}\leq 3/8)\\
& = \frac{m}{2}.
\end{align*}
Consequently, if we let $T(m)$ denote the maximum number of comparisons required to find a MSF in a graph with $m$ edges, we obtain the recurrence
\[
T(m) \leq \sum_{i} T(m_i) + O(m), \qquad \text{where }\sum_i m_i\leq m/2.
\]
This yields $T(m)=O(m)$, and completes the proof.
\end{proof}

\bibliographystyle{alpha}
\bibliography{ref}

\end{document}